\documentclass[letterpaper, 10 pt, conference]{ieeeconf}  % Comment this line out if you need a4paper

\IEEEoverridecommandlockouts                              % This command is only needed if 
\usepackage{graphicx} % Required for inserting images
\usepackage[font=footnotesize]{caption}
\usepackage{subcaption}
\usepackage{tikz}
\usepackage{diagbox}
\usepackage{xcolor}
\usepackage{booktabs}
\usepackage{longtable}
\usepackage{multirow}
\usepackage{multicol}
\usepackage{cite}
\usepackage{lipsum}  
\usepackage{amssymb}
\usepackage{mathtools}

\usepackage{color}
\usepackage{ifthen}
\newboolean{com}
\setboolean{com}{false}% hide com

\usepackage{algorithm} 
\usepackage{algpseudocode}

\newtheorem{ex}{Example}
\newtheorem{rem}{Remark}

\title{\LARGE \bf
Sparse Identification of Nonlinear Dynamics with Side Information (SINDy-SI)$^{*}$
}

\author{Gabriel F. Machado and Morgan Jones$^{1}$% <-this % stops a space
\thanks{*The first author of this work was sponsored by the Department of Automatic
Control and Systems Engineering of The University of Sheffield.}% <-this % stops a space
\thanks{$^{1}$ Gabriel F. Machado and Morgan Jones are with the Department of Automatic
Control and Systems Engineering, The University of Sheffield, Amy Johnson Building, Mappin Street, Sheffield, S1 3JD, UK.
        {\tt\small gfmachado1@sheffield.ac.uk, morgan.jones@sheffield.ac.uk}}%
}

\begin{document}

\maketitle
\thispagestyle{empty}
\pagestyle{empty}

%%%%%%%%%%%%%%%%%%%%%%%%%%%%%%%%%%%%%%%%%%%%%%%%%%%%%%%%%%%%%%%%%%%%%%%%%%%%%%%%
\begin{abstract}

Modern societies have an abundance of data yet good system models are rare. Unfortunately, many of the current system identification and machine learning techniques fail to generalize outside of the training set, producing models that violate basic physical laws. This work proposes a novel method for the Sparse Identification of Nonlinear Dynamics with Side Information (SINDy-SI). SINDy-SI is an iterative method that uses Sum-of-Squares (SOS) programming to learn optimally fitted models while guaranteeing that the learned model satisfies side information, such as symmetries and physical laws. Guided by the principle of Occam's razor, that the simplest or most regularized best fitted model is typically the superior choice, during each iteration SINDy-SI prunes the basis functions associated with small coefficients, yielding a sparse dynamical model upon termination. Through several numerical experiments we will show how the combination of side information constraints and sparse polynomial representation cultivates dynamical models that obey known physical laws while displaying impressive generalized performance beyond the training set.

\end{abstract}
\vspace{-0.2cm}
\section{Introduction}
System identification is the art of learning and building a mathematical model that represents reality. It is the hidden foundation of industrial society, being the first step taken in a control systems engineering approach for ensuring the desired behavior of both mechanical and electrical machines. For controller synthesis in safety critical situations, models must reliably generalize outside of the training data set. Unfortunately, many existing nonlinear system identification methods, such as the Nonlinear Auto-Regressive Moving Average model with eXogenous inputs (NARMAX) \cite{billings2013nonlinear}, the Sparse Identification of Nonlinear Dynamics (SINDy) \cite{brunton2016}, the Ho-Kalman algorithm \cite{ho1966effective}, the dynamic mode decomposition (DMD) algorithm \cite{schmid2010} as well as the Fuzzy modeling \cite{babuvska1996overview}, do not enforce physical constraints and hence produce unreliable models that violate basic physical laws.

One of the current central focuses of the Machine Learning (ML) community is physics informed learning, which seeks to integrate physical knowledge, such as conservation laws, symmetries and invariance, into data modeling \cite{karniadakis2021physics}. A prime example of this is the work of \cite{baddoo2023physics} which extends the DMD algorithm to accommodate structural matrix constraints while falling short in capturing nonlinear models. Another illustrative case lies in the realm of designing Neural Networks (NNs) with the capacity to encapsulate symmetries and uphold conservation laws within their architectural framework, as exemplified by \cite{mattheakis2019physical}. Additionally, researchers have explored the integration of domain knowledge into the development of NN loss functions, as demonstrated by \cite{atwya2022structure}, and the establishment of performance metrics for NN cross-validation, as seen in \cite{pannell2022physics}. Nevertheless, it's important to acknowledge that fitting a NN to data inherently entails a non-convex optimization problem, making it challenging to attain globally optimal solutions in general. Furthermore, these studies leave room for addressing a broader range of physical laws, such as invariant sets, monotonicity, and known equilibria, as potential constraints within the NN architecture. {However, notably~\cite{otto2023unified} presents a framework for the training phase of a model that comprises of the discovery and enforcement of specific system property of symmetry that extends to different ML models, including NNs.}
	
A promising approach to incorporate general physical knowledge into learned models has been to use Sum-of-Squares (SOS) programming to formulate constrained regression problems as convex Semi-Definite Programming (SDP) problems \cite{pitarch2019systematic}. This approach has also been applied to system identification in \cite{ahmadi2023}. It was noted in \cite{ahmadi2023} that sets of functions defined by polynomial constraints, which they term ``Side Information (SI)", a terminology adopted in this paper, offer a versatile means of expressing a wide spectra of physical knowledge. {SI is particularly useful in the identification of models in scarce data scenarios}. {This prior knowledge constraints are useful to many applications in system identification\cite{khosravi2021nonlinear,khosravi2023kernel} and control systems design \cite{luppi2023data}.} Nonetheless, these methods tend to overlook the fact that most physical systems can be characterized by relatively simple models, involving only a few relevant terms that govern the dynamics. Therefore these polynomial methods are prone to overfitting.

Occam's razor, a fundamental principle in scientific reasoning, implies that simpler explanations should be favored. In the context of system modeling, SINDy \cite{brunton2016} exemplifies this principle. SINDy  is a method that learns system models by iteratively solving Ordinary Least Squares (OLS) problems and removing basis functions corresponding to parameters below a user defined threshold. SINDy has been successfully applied to practical applications from Li-ion battery modeling~\cite{rodriguez2023discovering,ahmadzadeh2023physics} to discovering chemical reaction dynamics \cite{hoffmann2019reactive}. Numerous modifications of SINDy have been proposed, such as SINDy-PDE \cite{rudy2017data}, SINDy-c \cite{brunton2016sparse}, implicit-SINDy \cite{mangan2016inferring}, abrupt-SINDy \cite{quade2018sparse}, Lagrangian-SINDy \cite{chu2020discovering}. 

{There are only a few examples in the literature of combining SINDy with constrained contextual knowledge~\cite{loiseau2018constrained,champion2020unified,kaptanoglu2021physics,guan2021sparse}. In all of these examples linear constraint equations are added to the OLS problem. In this paper we extend these methods from linear constraints to SOS constraints, that allow for the enforcement of a wider class of contextual model knowledge. Our method unifies the approaches of using SOS side information constraints \cite{pitarch2019systematic,ahmadi2020} with the SINDy \cite{brunton2016} framework to devise the Sparse Identification of Nonlinear Dynamics with Side Information (SINDy-SI) algorithm. This algorithm iteratively tackles a sequence of SOS programs, simultaneously constraining side information while fitting trajectory data, and prunes basis functions during each iteration to yield a sparse model upon termination.}

% Like the original SINDy, SINDy-SI parameterizes the system model through a linear combination of nonlinear basis functions and iteratively tackles a sequence of optimization problems. However, unlike SINDy, which iteratively solves OLS problems, SINDy-SI iteratively solves SOS problems that incorporate side information constraints. During each iteration basis functions associated with small parameters are pruned from the basis library to yield a sparse model upon termination. Through cross-validation, we demonstrate that SINDy-SI outperforms both traditional SINDy and the SOS-based approach presented in \cite{ahmadi2023}.
\vspace{-0.2cm}
\subsection*{\textbf{Notation}}
\vspace{-0.2cm}
\noindent We define $\mathbb{N}$, $\mathbb{R}$, $\mathbb{R}_{\geq 0}$, and $\mathbb{R}_{> 0}$ as the set of natural, real, nonnegative real, and positive real numbers, respectively. We denote $\mathbb{R}^{m,n}$ as the set of real $m$-by-$n$ matrices and $\mathbb{S}^{n}$ as the set of symmetric matrices. Given $\mathbf{A} \in \mathbb{R}^{m,n}$, we denote $a_{i,j} \in \mathbb{R}$, $i,j \in \mathbb{N}$, as the $(i,j)$-th entry of $\mathbf{A}$. We define $\text{tr} \left[\mathbf{A}\right] = \sum_{i=1}^{n} a_{i,i}$ for $\mathbf{A} \in \mathbb{R}^{n,n}$. The norms $\Vert \mathbf{A} \Vert_{\text{F}} = \left( \text{tr} \left[\mathbf{A}\mathbf{A}^{\top}\right]\right)^{1/2}$, ${||\mathbf{A}||_1 = \sum_{i=1}^{m} \sum_{j=1}^{n} |a_{i,j}|}$, and the pseudo-norm $\Vert \mathbf{A} \Vert_{0} = \lim_{p \to 0^{+}} \left(\sum_{i=1}^{m} \sum_{j=1}^{n} \vert a_{i,j}\vert^{p} \right)^{1/p}$ are defined for $\mathbf{A} \in \mathbb{R}^{m,n}$. We denote $\mathcal{C}^{1}\left(X;Y\right)$ as the set of continuously-differentiable functions that maps $X$ to $Y$. {For $\mathbf{f} \in \mathcal{C}^{1}\left(X; Y \right)$ and $\mathbf{x} \in X$ we denote $f_i(\mathbf{x}) \in Y$ to mean the $i$'th component of the vector output of $\mathbf{f}$ evaluated at $\mathbf{x} \in X$.} For a vector $\mathbf{x} \in \mathbb{R}^{n}$, we define $\Vert \mathbf{x} \Vert_{2} = \left(\sum_{i=1}^{n} \vert x_{i}\vert^{2}\right)^{1/2}$. We denote $\mathcal{P}_d\left(X;\mathbb{R}^{n}\right)$, $d \in \mathbb{N}$, as the set of $d$-degree polynomial vector-valued functions. We denote $\Sigma \left( X;\mathbb{R} \right)$ as the set of Sum-of-Squares polynomials, that is, $p \in \Sigma \left( X;\mathbb{R} \right)$ if there exist $\{q_{i}\}_{i=1}^k \in \mathcal{P}_{d} \left( X;\mathbb{R} \right)$ such that $p(x) = \sum_{i=1}^{k} q_{i}(x)^{2}$. We denote $\mathcal{N}\left(\mu,\sigma,X\right)$ as the set of numbers from the normal distribution with mean $\mu \in X$ and standard deviation $\sigma \in \mathbb{R}_{\geq 0}$.
\vspace{-0.2cm}
\section{Methods for Nonlinear System Identification}\label{sec:prob-form}

Consider a dynamical system represented by the system of autonomous ordinary differential equations (ODEs) in an initial value problem (IVP):
\vspace{-0.2cm}
\begin{equation}
    \dot{\mathbf{x}}(t) = \mathbf{f}(\mathbf{x}(t)),\hspace{0.5cm} \mathbf{x}(0) = \mathbf{x}_{0},\hspace{0.5cm}t \in \mathbb{R}_{\geq 0}
    \label{eq:ivp}
\end{equation}
where $\mathbf{f} \in \mathcal{C}^{1}\left(\mathbb{R}^{n};\mathbb{R}^{n}\right)$ is the vector field and $\mathbf{x}_{0} \in \mathbb{R}^{n}$ the initial condition. Throughout this paper we will make the assumption that Equation~\eqref{eq:ivp} is well defined, that is, there exists a unique solution $\mathbf{s} \in \mathcal{C}^{1}\left(\mathbb{R}_{\geq 0} \times \mathbb{R}^{n};\mathbb{R}^{n}\right)$ that satisfies the IVP; i.e., $\dot{\mathbf{s}}(t,\mathbf{x}_{0}) = \mathbf{f}\left(\mathbf{s}(t,\mathbf{x}_{0})\right)$ and $\mathbf{s}(0,\mathbf{x}_{0}) = \mathbf{x}_{0}$.

Suppose that the vector field, $\mathbf{f}: \mathbb{R}^{n} \to \mathbb{R}^{n}$ from IVP~\eqref{eq:ivp} is unknown but we have access to trajectory data snapshots $D=\{\mathbf{s}(t_i,\mathbf{x}_j)\}_{1 \le i \le r,1 \le j \le m}$, where {the number of sampled timesteps is} $r \in \mathbb{N}$ and {the number of sampled trajectories with different intial conditions is} $m \in \mathbb{N}$. The goal of this paper is to estimate $\hat{\mathbf{f}} \approx \mathbf{f}$ from the given data $D$. To address this problem, we first formulate it as a supervised learning problem. Noting that
\vspace{-0.2cm}
\begin{align} \label{eq: approx diff}
    \mathbf{f}(\mathbf{s}(t_i,\mathbf{x}_j)) \approx \frac{\mathbf{s}(t_i,\mathbf{x}_j) - \mathbf{s}(t_{i-1},\mathbf{x}_j)}{t_i -t_{i-1}},
\end{align}
we expand our data set to include both inputs and outputs of the unknown vector field, $D=\{\mathbf{s}(t_i,\mathbf{x}_j),\mathbf{f}(\mathbf{s}(t_i,\mathbf{x}_j))\}_{i,j}$. {Hence, the data set comprehends trajectories with initial conditions $\mathbf{x}_j$ and vector field outputs $\mathbf{f}(\mathbf{s}(t_i,\mathbf{x}_j))$ sampled at timestamps $t_i$.} To account for the fact that we do not have error free access to $\{\mathbf{f}(\mathbf{s}(t_i,\mathbf{x}_j))\}_{i,j}$, noise will be injected into the data set {through Eq.~\eqref{eq: data noise injection}} before proceed with numerical examples. This noise represents both measurement error and errors accrued through the approximate differentiation in Eq.~\eqref{eq: approx diff}.

The problem of estimating the vector field can then be formulated as the following optimization problem,
\vspace{-0.2cm}
\begin{align} \label{opt: fitting a vector field}
    \min_{\hat{\mathbf{f}}} \sum_{i,j} \left\Vert \mathbf{f}(\mathbf{s}(t_i,\mathbf{x}_j))- \hat{\mathbf{f}}(\mathbf{s}(t_i,\mathbf{x}_j)) \right\Vert_2^2.
\end{align}

\vspace{-0.2cm}
\subsection{The OLS Approach to System Identification}
Optimization Problem~\eqref{opt: fitting a vector field} is an infinite dimensional problem, having a decision variable that is a function. To make this problem tractable we parameterize the decision variable as a finite vector of coefficients. More specifically, we seek to learn models of the following form,
\vspace{-0.2cm}
\begin{equation}
    \begin{split}
        \hat{\mathbf{f}}(\mathbf{x}) = \mathbf{W}^{ \top} \boldsymbol{\phi}(\mathbf{x}),
    \end{split}
    \label{eq:poly-model}
\end{equation}
where $\mathbf{W} \in \mathbb{R}^{h,n}$ is a matrix of coefficients, $\boldsymbol{\phi}: \mathbb{R}^n \to \mathbb{R}^{h}$. {Here, $h \in \mathbb{N}$ represents the number of basis elements in $\hat{\mathbf{f}}(\mathbf{x})$. The larger $h$ is the better $\hat{\mathbf{f}}$ will fit the data.}

When considering models given by Eq.~\eqref{eq:poly-model},  Opt.~\eqref{opt: fitting a vector field} reduces to the following Ordinary Least Squares (OLS) problem,
\vspace{-0.2cm}
\begin{equation}
    \begin{split}
       \mathbf{W}^* \in \arg \min_{\mathbf{W} \in \mathbb{R}^{h,n}} \left\Vert \mathbf{Y} - \boldsymbol{\Phi} \mathbf{W} \right\Vert_{\text{F}}^{2},\\
    \end{split}
    \tag{OLS}
    \label{eq:op-ols}
\end{equation}
where $\mathbf{Y} \in \mathbb{R}^{m r, n}$ and $\boldsymbol{\Phi} \in \mathbb{R}^{m r, h}$ are defined by: \vspace{-0.2cm}
\begin{equation} \label{eq: Y}
    \begin{split}
        \mathbf{Y} = \begin{bmatrix}
            \mathbf{F}_{1}^{\top} & \hdots &
            \mathbf{F}_{m}^{\top}
        \end{bmatrix}^{\top},~&
        \boldsymbol{\Phi} = \begin{bmatrix}
            \mathbf{P}_{1}^{\top} &
            \hdots &
            \mathbf{P}_{m}^{\top}
        \end{bmatrix}^{\top},
    \end{split}
\end{equation}
and where matrices $\mathbf{F}_{j}$ and $\mathbf{P}_{j}$ are constructed as \vspace{-0.2cm}
\begin{equation}
    \begin{split}
        \mathbf{F}_{j} &= \begin{bmatrix}
            f_{1}(\mathbf{s}(t_{1},\mathbf{x}_{j})) & \cdots & f_{n}(\mathbf{s}(t_{1},\mathbf{x}_{j}))\\
            \vdots & \vdots & \vdots\\
            f_{1}(\mathbf{s}(t_{r},\mathbf{x}_{j})) & \cdots & f_{n}(\mathbf{s}(t_{r},\mathbf{x}_{j}))\\
        \end{bmatrix} \in \mathbb{R}^{r,n},\\
        \mathbf{P}_{j} &= \begin{bmatrix}
            \phi_{1}(\mathbf{s}(t_{1},\mathbf{x}_{j})) & \cdots & \phi_{h}(\mathbf{s}(t_{1},\mathbf{x}_{j}))\\
            \vdots & \vdots & \vdots\\
            \phi_{1}(\mathbf{s}(t_{r},\mathbf{x}_{j})) & \cdots & \phi_{h}(\mathbf{s}(t_{r},\mathbf{x}_{j}))\\
        \end{bmatrix} \in \mathbb{R}^{r,h}.\\
    \end{split}
    \label{eq:matrices_Fj_Pj}
\end{equation}

% \mj{[I've deleted the definition of F and S. Maybe you can work it back in or define it closer to the cross validation stuff.]}

It is well known that if  the matrix $\boldsymbol{\Phi}^{\top} \boldsymbol{\Phi}$ is non-singular the~\eqref{eq:op-ols} optimization problem has the following closed-form solution:
\vspace{-0.4cm}
\begin{equation}
    \begin{split}
        \mathbf{W}^{*} = (\boldsymbol{\Phi}^{\top} \boldsymbol{\Phi})^{-1} \boldsymbol{\Phi}^{\top} \mathbf{Y}.\\
    \end{split}
    \label{eq:ols-solution}
\end{equation}

\vspace{-0.2cm}
\subsection{Sparse Identification of Nonlinear Dyamics (SINDy)}

By choosing our vector-valued basis functions $\boldsymbol{\phi}$ in \eqref{eq:poly-model} wisely, for instance, opting for universal predictors like monomial basis functions, radial basis functions, and similar options, we unlock the capability to approximate any continuous vector field to arbitrary precision. Nevertheless, a heedless expansion of this basis function set could lead to a model that becomes excessively sensitive to training data errors, ultimately hindering its ability to generalize effectively. To prevent overfitting, we regularize the objective in \eqref{eq:op-ols}, \vspace{-0.2cm}
\begin{equation}
    \begin{split}
       \mathbf{W}^* \in \arg \min_{\mathbf{W} \in \mathbb{R}^{h,n}} \left\Vert \mathbf{Y} - \boldsymbol{\Phi} \mathbf{W} \right\Vert_{\text{F}}^{2} + \xi||\mathbf{W}||_0,\\
    \end{split}
    \label{eq:op-ols L0}
\end{equation}
{where the parameter $\xi \in \mathbb{R}_{\geq 0}$ weights sparseness and can be selected using standard machine learning techniques like cross-validation and hyperparameter tuning. The $\ell_0$-pseudonorm makes the objective function in Opt.~\eqref{eq:op-ols L0} non-convex and discontinuous. Since the size of this problem grows with respect to the size of $\mathbf{W}$, it follows that finding a global minimizer $\mathbf{W}^*$ becomes computationally expensive. To address this issue, approximation methods are usually employed such as heuristics, convex relaxations or proximal projections as in \cite{ghayem2018sparse}.}

{SINDy \cite{brunton2016} is a heuristic algorithm that solves Opt. \eqref{eq:op-ols L0} and is summarized in Algorithm \ref{alg:sindy}. SINDy works by solving \eqref{eq:op-ols} iteratively and removing basis functions that correspond to values less than a user-defined threshold. In a more formal manner, the \eqref{eq:op-ols} problem is solved first and its solution, $\mathbf{W}^{*}_{1}$, is then input into Algorithm \ref{alg:sparse}. This algorithm is a thresholding algorithm that collects all the elements of $\mathbf{W}^{*}_{1}$ that are below a threshold parameter $\lambda \in \mathbb{R}_{> 0}$, defined by the user or tuned as a hyperparameter, and stores their indexes in $E$. Following that, \eqref{eq:op-ols} is solved a second time alongside the equality constraints $w_{i,j} = 0$ for all $(i,j) \in E$:}
% SINDy \cite{brunton2016} is a heuristic algorithm for solving Opt.~\eqref{eq:op-ols L0} and is summarized in Algorithm~\ref{alg:sindy}. SINDy works by iteratively solving \eqref{eq:op-ols} and removing basis functions that correspond to values bellow a user defined threshold. More formally, first the \eqref{eq:op-ols} problem is solved and its solution, $\mathbf{W}^*$, is inputted into Algorithm~\ref{alg:sparse}. Algorithm~\ref{alg:sparse} is a thresholding algorithm that collects all the elements of $\mathbf{W}^*$ that are bellow a threshold parameter $\lambda \in \mathbb{R}_{> 0}$, defined by the user or tuned as a hyperparameter, and stores the indexes of these parameters in $E$. Next, \eqref{eq:op-ols} is solved again with the constraints $w_{i,j} = 0$ for all $(i,j) \in E$:
\vspace{-0.2cm}
\begin{equation}
    \begin{split}
         \mathbf{W}^* \in \arg \min_{\mathbf{W} \in \mathbb{R}^{h,n}} \left\Vert \mathbf{Y} - \boldsymbol{\Phi} \mathbf{W} \right\Vert_{\text{F}}^{2}&\\
        \text{subject to:} \quad 
        w_{i,j} = 0,~\text{for all}~(i,j) \in E.&\\
    \end{split}
    \label{eq:op-sindy}
\end{equation}

\begin{algorithm}[htb]
	\caption{Sparseness algorithm}
	\begin{algorithmic}[1]
        \State \textbf{Inputs:} $\mathbf{W}^{*} \in \mathbb{R}^{h,n}$, $\lambda \in \mathbb{R}_{> 0}$
        \State Set $E = \emptyset \subset \mathbb{N} \times \mathbb{N}$
        \For {$i = 1, \dots, h$}
            \For {$j = 1, \dots, n$}
                \If {$\vert w_{i,j}^{*} \vert < \lambda$}
                    \State $E = E \cup \left\{(i,j)\right\}$
                \EndIf
            \EndFor
        \EndFor
        \State \textbf{Output:} $E$
	\end{algorithmic}
    \label{alg:sparse}
\end{algorithm}
% \vspace{-0.2cm}
\begin{algorithm}[htb]
	\caption{SINDy}
	\begin{algorithmic}[1]
        \State Solve \eqref{eq:op-ols} to obtain $\mathbf{W}^{*}_{1} \in \mathbb{R}^{h,n}$
        \State Set $N \in \mathbb{N}$ and $\lambda \in \mathbb{R}_{> 0}$
        \For{$i = 2, \dots, N$}
            \State Run Algorithm \ref{alg:sparse} with inputs: $\mathbf{W}^{*}_{i-1}$ and $\lambda$
            \If{$E = \emptyset$}
                \State \textbf{Output:} $\mathbf{W}^{*}_{i-1}$
            \EndIf
            \State {Solve \eqref{eq:op-sindy} to obtain $\mathbf{W}^{*}_{i} \in \mathbb{R}^{h,n}$}
        \EndFor
	\end{algorithmic}
    \label{alg:sindy}
\end{algorithm}
\vspace{-0.20cm}
{
\begin{rem} \label{rem: non unique ols}
    Because the SINDy algorithm iteratively solves \eqref{eq:op-ols}, the size of the basis function $\boldsymbol{\phi}$ must be balanced. In a situation of more basis elements than data, the solution $\mathbf{W}^*$ in Equation \eqref{eq:ols-solution} cannot be determined uniquely. In this case, the regressor matrix $\boldsymbol{\Phi}$ has not full rank, and \eqref{eq:op-ols} becomes a system of ``underdetermined" equations with infinitely many solutions. As a result, with scarce data, SINDy may underperform and inadvertently remove important basis functions that are critical in describing the true dynamics.
\end{rem}

As we will see in the next subsection, tightening \eqref{eq:op-ols} by introducing constraints can help ease non-uniqueness challenges for scarce data sets. } \vspace{-0.25cm}

\subsection{Identification with Side Information Constraints}

In various practical scenarios, we possess contextual knowledge pertaining to system dynamics beyond just trajectory data. Take, for example, population dynamics, where we are aware that a population's minimum value is zero, or in the case of physical systems, where we understand the principle of momentum conservation. {This prior knowledge is supplementary data referred to as ``side information" and it plays a crucial role, especially when data availability is limited.} Enforcing our model to adhere to this side information can prove instrumental in alleviating the challenges posed by data scarcity.

We can pose such side information as polynomial constraints. For example, if it is known that the true dynamics have the following symmetry $\mathbf{f}(\mathbf{x}) = -\mathbf{f}(-\mathbf{x})$, then we can enforce our learned model to satisfy $\hat{\mathbf{f}} \in S = \left\{\mathbf{f} \in \mathcal{C}^{1}\left(\mathbb{R}^{n}:\mathbb{R}^{n}\right): \mathbf{f}(\mathbf{x}) = -\mathbf{f}(-\mathbf{x})\right\}$. There are other types of side information, such as monotonicity $\{f_{i}(\mathbf{x})\}_{i=1}^{n} \geq 0$, that involve inequalities and can be enforced in the model as $\hat{\mathbf{f}} \in S = \left\{\mathbf{f} \in \mathcal{C}^{1}\left(\mathbb{R}^{n}:\mathbb{R}^{n}\right): \{f_{i}(\mathbf{x})\}_{i=1}^{n} \geq 0\right\}$, for example. {To ensure that side information constraints are numerically tractable, unlike SINDy we now only consider polynomial models for the following reasons: 1) For sufficiently large degree $d$, they can approximate any continuous function over a compact set to arbitrary precision, according to the Stone-Weierstrass approximation theorem; 2) Sum-of-Squares (SOS) programming can efficiently constrain side information that involve functional inequalities (such as monotonicity) \cite{ahmadi2023}.}

We now modify \eqref{eq:op-ols} to include given side information sets $\{S_{i}\}_{i=1}^{q}$ and polynomial constraints, leading us to the following optimization problem:
\vspace{-0.2cm}
\begin{align}    \label{eq:op-side-info} \tag{SI}
         &\mathbf{W}^* \in \arg \min_{\mathbf{W} \in \mathbb{R}^{h,n}} \left\Vert \mathbf{Y} - \boldsymbol{\Phi} \mathbf{W} \right\Vert_{\text{F}}^{2}  \\ \nonumber
       & \text{subject to: }
        \hat{\mathbf{f}}(\mathbf{x}) = \mathbf{W}^{\top} \boldsymbol{\phi}(\mathbf{x}),~\hat{\mathbf{f}} \in \mathcal{P}_{d}\left(\mathbb{R}^{n};\mathbb{R}^{n}\right) \cap \{S_{i}\}_{i=1}^q.
\end{align}
{We use standard techniques to lift \eqref{eq:op-side-info} to a SDP problem in order to solve it. A relaxation of this optimization problem is done by searching for a matrix $\mathbf{M} \in \mathbb{S}^{n}$ and using the Schur complement:}
% To solve \eqref{eq:op-side-info} we use standard techniques to lift this problem to an SDP. The optimization problem is relaxed as we search for a matrix $\mathbf{M} \in \mathbb{S}^{n}$ and use the Schur complement:
\vspace{-0.2cm}
\begin{equation}
\centering
\begin{split}
    \mathbf{M} &\succeq ( \mathbf{Y} - \boldsymbol{\Phi} \mathbf{W})^{\top}( \mathbf{Y} - \boldsymbol{\Phi} \mathbf{W})\\
    &\iff \begin{bmatrix}
        \mathbf{M} & ( \mathbf{Y} - \boldsymbol{\Phi} \mathbf{W})^{\top}\\
        ( \mathbf{Y} - \boldsymbol{\Phi} \mathbf{W}) & \mathbf{I}_{m r}
    \end{bmatrix} \succeq 0\\
    &\implies \text{tr} \left[ \mathbf{M} \right] \geq \text{tr} \left[ ( \mathbf{Y} - \boldsymbol{\Phi} \mathbf{W})^{\top}( \mathbf{Y} - \boldsymbol{\Phi} \mathbf{W}) \right].
\end{split}
\end{equation}
{An optimization problem with a linear objective function can be formulated using $\text{tr} \left[ \mathbf{M} \right]$:}
% Minimizing $\text{tr} \left[ \mathbf{M} \right]$ leads us to formulate the optimization problem with linear objective function:
\vspace{-0.2cm}
\begin{align}  \label{eq:op-sdp-side-info}
        & \min_{\mathbf{W} \in \mathbb{R}^{h,n}, \mathbf{M} \in \mathbb{S}^{n}} \text{tr} \left[ \mathbf{M} \right]\\ \nonumber
       & \text{subject to: }
        \hat{\mathbf{f}}(\mathbf{x}) = \mathbf{W}^{\top} \boldsymbol{\phi}(\mathbf{x}), \text{ }
        \hat{\mathbf{f}} \in \mathcal{P}_{d}\left(\mathbb{R}^{n};\mathbb{R}^{n}\right) \cap \{S_{i}\}_{i=1}^q,\\ \nonumber
        &\begin{bmatrix}
            \mathbf{M} & ( \mathbf{Y} - \boldsymbol{\Phi} \mathbf{W})^{\top}\\
            ( \mathbf{Y} - \boldsymbol{\Phi} \mathbf{W}) & \mathbf{I}_{m r}
        \end{bmatrix} \succeq 0.
\end{align}

\section{Sparse Identification of Nonlinear
Dynamics with Side Information (SINDy-SI)}\label{sec:methodology}

%\mj{[I'd like to have the (SI) problem with L0 reg and say we are hurestically trying to solve that here.]}

%\mj{[I think you have this opt in the previous section but probably need it here.]}

% \mj{[Maybe Quote theorem that polynomial can arbitrarily approx any continuous vector field over compact domain. However increasing d leads to an exponentially dense parameterization of the vector field. However, .... for some reason (SINDY paper argument) vector fields are sparse (Ocams razor principle). ]} 

% \mj{[I'm in two minds about quoting that theorem. Maybe it doesnt fit? It would make the paper more technical. But doesnt add to the contribution unless we could use it together with a SINDy convergence result.]}

{Although \eqref{eq:op-side-info} incorporates side information constraints, like \eqref{eq:op-ols}, it is still susceptible to overfitting due to the lack of regularization. In order to prevent this overparameterization, we regularize the objective function in \eqref{eq:op-side-info}:
\vspace{-0.1cm}
\begin{align}     \label{eq:op-side-info L0}
       &  \mathbf{W}^* \in \arg \min_{\mathbf{W} \in \mathbb{R}^{h,n}} \left\Vert \mathbf{Y} - \boldsymbol{\Phi} \mathbf{W} \right\Vert_{\text{F}}^{2}+ \xi_{1} ||\mathbf{W}||_0 + \xi_{2} ||\mathbf{W}||_1\\ \nonumber 
       &  \text{subject to: }
        \hat{\mathbf{f}}(\mathbf{x}) = \mathbf{W}^{\top} \boldsymbol{\phi}(\mathbf{x}),~\hat{\mathbf{f}} \in \mathcal{P}_{d}\left(\mathbb{R}^{n};\mathbb{R}^{n}\right) \cap \{S_{i}\}_{i=1}^q,
\end{align}}
where the parameters $\xi_{1},\xi_{2} \in \mathbb{R}_{\geq 0}$ weight sparseness. Both $\Vert \mathbf{W} \Vert_{0}$ and $\Vert \mathbf{W} \Vert_{1}$ make the problem computationally intractable. Therefore, we address this issues relaxing $\xi_{2} \Vert \mathbf{W} \Vert_{1}$ to linear inequalities $-\gamma \leq \xi_{2} w_{i,j} \leq \gamma$, minimizing over $\gamma$, and a heuristic algorithm to minimize $\xi_{1} \Vert \mathbf{W} \Vert_{0}$ in Opt. \eqref{eq:op-side-info L0}. This approach is called SINDy-SI (Alg. \ref{alg:sparse-side-info}). Thus, SINDy-SI iteratively reduces the basis functions to yield a simple polynomial structure while satisfying side information constraints. More specifically, SINDy-SI solves a sequence of SOS programs. During each iteration, Algorithm~\ref{alg:sparse} is executed and the indexes of the polynomial coefficients that fall bellow a user defined threshold are collected into the set $E$. Then on the next iteration these coefficients are constrained to be zero and the following optimization problem is solved:{
%\vspace{-0.2cm}
\begin{align}    \label{eq:op-sdp-side-info-sparse}
         & \min_{\mathbf{W} \in \mathbb{R}^{h,n}, \mathbf{M} \in \mathbb{S}^{n},\gamma \in \mathbb{R}_{\geq 0}} \text{tr} \left[ \mathbf{M} \right] + \gamma \\ \nonumber
        & \text{subject to: } \hat{\mathbf{f}}(\mathbf{x}) = \mathbf{W}^{\top} \boldsymbol{\phi}(\mathbf{x}),~
        \hat{\mathbf{f}} \in \mathcal{P}_{d}\left(\mathbb{R}^{n};\mathbb{R}^{n}\right) \cap \{S_{i}\}_{i=1}^q,
        \\ \nonumber & w_{i,j} = 0 \text{ for}~ (i,j) \in E, \quad  -\gamma \leq \xi_{2} w_{i,j} \leq \gamma \text{ for all } (i,j),
        \\ \nonumber 
       & \begin{bmatrix}
            \mathbf{M} & ( \mathbf{Y} - \boldsymbol{\Phi} \mathbf{W})^{\top}\\
            ( \mathbf{Y} - \boldsymbol{\Phi} \mathbf{W}) & \mathbf{I}_{m r}
        \end{bmatrix}\succeq 0.%, \{-\gamma \leq \xi_{2} w_{i,j} \leq \gamma \}_{i,j=1}^{h,n}.
\end{align}}

\begin{algorithm}[htb]
	\caption{SINDy-SI}
	\begin{algorithmic}[1]
 {
        \State Solve \eqref{eq:op-side-info} to obtain $\mathbf{W}^{*}_{1} \in \mathbb{R}^{h,n}$
        \State Set $N \in \mathbb{N}$, $\lambda \in \mathbb{R}_{> 0}$, $\xi_2 \in \mathbb{R}_{\geq 0}$
        \For{$i = 2, \dots, N$}
            \State Run Algorithm \ref{alg:sparse} with inputs: $\mathbf{W}^{*}_{i-1}$ and $\lambda$
            \If{$E = \emptyset$}
                \State \textbf{Output:} $\mathbf{W}^{*}_{i-1}$
            \EndIf
            \State Solve \eqref{eq:op-sdp-side-info-sparse} to obtain $\mathbf{W}^{*}_{i} \in \mathbb{R}^{h,n}$
        \EndFor}
	\end{algorithmic}
    \label{alg:sparse-side-info}
\end{algorithm}
%\vspace{-0.25cm}
\section{Numerical results}\label{sec:simulations}
{All numerical examples presented in this section take place in the scenario of limited training data, specifically when $h > r$. In such instances it is expected that methods involving the \eqref{eq:op-ols} problem, like SINDy, will exhibit poor performance, see Remark~\ref{rem: non unique ols}. Our goal is to demonstrate the useful of SINDy-SI in such scarce data scenarios. We next detail the implementation and performance evaluation before proceeding with the numerical examples.}

% \gm{I can separate the Examples in two: 1) Comparison to OLS, SINDy, and SI for the Lorenz system. 2) SINDy-SI alone for a nonpolynomial system and comparison with Taylor expansion and other regularization methods like $L_{1}$ (and possibly $L_{2}$).

% In 1), for Lorenz system, we have two cases: a) "Low" noise in data $S$ and $Y$. b) "High" noise in data $S$ and $Y$. Show for each case a percentage of how SINDy-SI performs better (or not) wrt the other approaches $50$ folds are okay but $100$ would be better, I'll see how long this takes to run. Cases a) and b) will have different trajectories.

% In 2), for a nonpolynomial system, I can show the mean and variance on the coefficients of the model in all folds.}
\paragraph{Data Generation and Processing}
{To create a data set $D$, we collect a total of $m \in \mathbb{N}$ trajectories with $r \in \mathbb{N}$ timestamps {by simulating the ODEs using Matlab's inbuilt \texttt{ODE45} function.} Although vector field outputs are assumed to be measurable, no error free measurements are considered as we inject noise into our data set as follows,
\begin{equation}
    \begin{split}
        D=\{\mathbf{s}(t_i,\mathbf{x}_j)+\epsilon_{i,j},\mathbf{f}(\mathbf{s}(t_i,\mathbf{x}_j))+\eta_{i,j}\}_{i,j=1}^{r,m},
    \end{split}
    \label{eq: data noise injection}
\end{equation}
where $\boldsymbol{\epsilon} \in \mathcal{N}(\mathbf{0},\sigma_{\mathbf{S}},\mathbb{R}^{m r,n})$, $\boldsymbol{\eta} \in \mathcal{N}(\mathbf{0},\sigma_{\mathbf{Y}},\mathbb{R}^{m r,n})$. {Before learning any model from this generated data we normalize the regressor matrix, $\mathbf{P}_j$, from Eq.~\eqref{eq:matrices_Fj_Pj} to make it have unitary columns under the $\ell_{2}$-norm. Without this normalization it would not be sensible to use a single $\lambda>0$ threshold in SINDy or SINDy-SI since small coefficients could still have a large impact on the model output if they're associated with un-normalized basis functions that take large values.   }

\paragraph{Cross Validation}
The power of SINDy-SI lies in practical situations where
we have limited noisy data. To demonstrate this ability we compare SINDy-SI to several other methods based on a leave ($m-1$)-out cross-validation technique, which uses a single fold to train the model and all the remaining folds to test the model. More specifically, we split the data set $D$ into $m$ folds, each fold corresponding to a single trajectory.  We define the cost $J_k$ of training on fold $k \in \mathbb{N}$, $1 \leq k \leq m$, and testing on the remaining folds as %\vspace{-0.25cm}
\begin{equation}
    \begin{split}
    J_{k} = \frac{\underset{j \neq k}{\sum_{i=1}^{r} \sum_{j=1}^{m}} \left\Vert \mathbf{f}(\mathbf{s}(t_i,\mathbf{x}_{j})) - \hat{\mathbf{f}}\left(\mathbf{s}(t_i,\mathbf{x}_{j})\right) \right\Vert_{2}}{(m-1) r}
    \end{split}
    \label{eq:cost}
\end{equation}
such that $(\mathbf{s}(t_i,\mathbf{x}_{j}),\mathbf{f}(\mathbf{s}(t_i,\mathbf{x}_{j}))\in\left(\mathbf{S}_{j},\mathbf{F}_{j}\right)$ with $\mathbf{S}_j$ representing trajectory data %\vspace{-0.25cm}
\begin{equation}
    \begin{split}
        \mathbf{S}_{j} = \begin{bmatrix}
            s_{1}(t_{1},\mathbf{x}_{j}) & \cdots & s_{n}(t_{1},\mathbf{x}_{j})\\
            \vdots & \vdots & \vdots\\
            s_{1}(t_{r},\mathbf{x}_{j}) & \cdots & s_{n}(t_{r},\mathbf{x}_{j})\\
        \end{bmatrix} \in \mathbb{R}^{r,n},
    \end{split}
\end{equation}
whereas $\mathbf{F}_j$, as in Eq. \eqref{eq:matrices_Fj_Pj}, represents the vector field outputs. 

\paragraph{SINDy-SI Implementation}
To fairly compare and benchmark SINDy-SI with SINDy we use a fixed handpicked threshold, $\lambda>0$. However, in practice $k$-fold cross validation hyperparameter tuning should be employed, which may result in differing optimal thresholds for each method. The other SINDy-SI parameter is selected as $N = h n$, since it follows from Theorem 2.1 in \cite{sindy2019convergence} that $N = h n$ is the upper bound on the number of possible eliminations from $E$ outputted by Algorithm \ref{alg:sparse}. However, for practical applications where computational time is limited it is advisable to select $N$ to be smaller than this upper bound. 

To enforce side information constraints, for simplicity, we work over some sufficiently large ball of radius $R>0$ centered at $\mathbf{c}_0 \in \mathbb{R}^n$ that bounds our normalized data set. Because this set is compact, the polynomial inequalities resulting from the side information constraints reduce to SOS constraints for sufficiently large degree by Putinar's \textit{Positivstellensatz} \cite{putinar1993}. Each resulting SOS problem is then solved using SOSTOOLS \cite{sostools} with SDP solver MOSEK \cite{mosek}.

}

\begin{ex}[The Lorenz system]
    Consider the Lorenz system \cite{lorenz1963,lorenz1989} with its usual parameters:
%\vspace{-0.2cm}
\begin{equation}
    \begin{split}
        \dot{\mathbf{x}} = \mathbf{f}(\mathbf{x}) = \begin{bmatrix}
            10(x_{2}-x_{1})\\
            28 x_{1} - x_{2} - x_{1} x_{3}\\
            -(8/3) x_{3} + x_{1} x_{2}
        \end{bmatrix}
    \end{split}
    \label{eq:lorenz-ode}
\end{equation}

{Data were generated from randomized initial conditions. We considered three different noise scenarios, each with the same $m = 100$ trajectories and $r = 40$ timestamps linearly spaced within $[0,10] $, but with different amounts of variation in the noise injected into the data according to Eq.~\eqref{eq: data noise injection}. The performance of different methods is presented in Table~\ref{tab:ex1} and Fig.~\ref{fig:lorenz-deg5}. All methods fitted a $d = 5$ degree polynomial model consisting of $h = 56$ monomial basis functions (note $h > r$). SINDy-SI and (SI) both enforced the following side information on the Lorenz system: 1) The origin as an equilibrium point $S_{1} = \{\mathbf{f} \in \mathcal{C}^{1}(\mathbb{R}^{n};\mathbb{R}^{n}): \mathbf{f}(\mathbf{x}) = \mathbf{0}\}$; 2) The vector field has coordinate directional monotonicity $S_{2} = \{\mathbf{f} \in \mathcal{C}^{1}(\mathbb{R}^{n};\mathbb{R}^{n}): \{-\partial f_{i}/\partial x_{i}(\mathbf{x})\}_{i=1}^{n} \geq 0\}$. The latter side information can be cast as SOS constraints of appropriate degree $\{-\partial \hat{f}_{i}/\partial x_{i}(\mathbf{x}) - z_{i}(\mathbf{x}) g(\mathbf{x})\}_{i=1}^{n} \in \Sigma\left(\mathbb{R}^{n};\mathbb{R}\right)$ with SOS multipliers $\{z_{i}(\mathbf{x})\}_{i=1}^{n} \in \Sigma\left(\mathbb{R}^{n};\mathbb{R}\right)$ and $g(\mathbf{x}) = 75^2 - \Vert \mathbf{x} - [0~0~25]^{\top}\Vert_{2}^2$. In Table~\ref{tab:ex1}, SINDy-SI is executed without $\ell_{1}$ regularization ($\xi_2=0$ in in Opt.~\eqref{eq:op-sdp-side-info-sparse}). Here, SINDy and SINDy-SI both used the threshold parameter $\lambda=0.1$.

To assess the effect of the inclusion of $\ell_{1}$ regularization ($\xi_2 \ne 0$ in Opt.~\eqref{eq:op-sdp-side-info-sparse}), in Figure \ref{fig:lorenz-deg5}, we present a comparison between \eqref{eq:op-side-info} and SINDy-SI variants. Figure \ref{fig:a} presents the ``low" noise scenario and shows that SINDy-SI has superior performance against \eqref{eq:op-side-info}. Moreover, the performance gap between SINDy-SI and \eqref{eq:op-side-info} is enlarged in Figure \ref{fig:b} while the third case with ``high" noise, depicted in Figure \ref{fig:c}, shows that SINDy-SI with $\lambda = 0.1$ and $\xi_{2} = 0.01$ substantially outperforms \eqref{eq:op-side-info}. From the results in Table \ref{tab:ex1} and Figure \ref{fig:lorenz-deg5}, we can see how SINDy-SI is able to provide valuable context to the model especially with scarce training data with high noise.}

% \gm{Figure \ref{fig:a} presents the ``low" noise scenario and shows SINDy-SI has superior performance against \eqref{eq:op-ols} and \eqref{eq:op-side-info} for all tests while it outperforms SINDy in $56\%$ of the tests. Moreover, the performance gap between SINDy-SI and SINDy is enlarged to $64\%$ of the tests in Figure \ref{fig:b} while in the third case with ``high" noise, presented in Figure \ref{fig:c}, shows SINDy-SI substantially outperforms SINDy in $94\%$ of the tests.}

%\vspace{-0.25cm}
\begin{figure*}
\raisebox{+5cm}
     \centering
     \begin{subfigure}{0.325\textwidth}
         \centering
         \includegraphics[width=\textwidth]{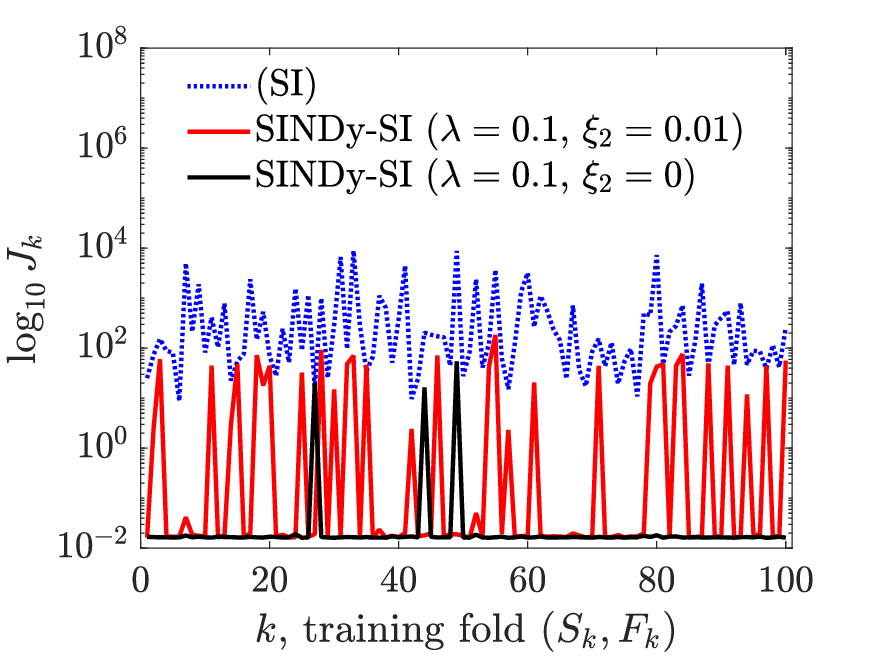}
         \caption{$\sigma_{\mathbf{S}} = 10^{-4}$ and $\sigma_{\mathbf{Y}} = 10^{-2}$}
         \label{fig:a}
     \end{subfigure}
     \hfill
     \begin{subfigure}{0.325\textwidth}
         \centering
         \includegraphics[width=\textwidth]{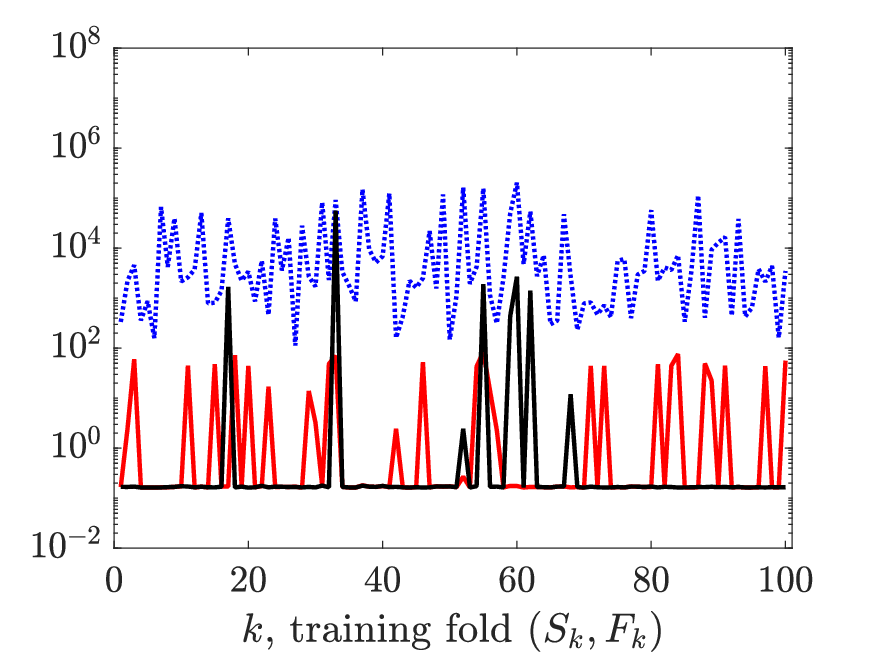}
         \caption{$\sigma_{\mathbf{S}} = 10^{-3}$ and $\sigma_{\mathbf{Y}} = 10^{-1}$}
         \label{fig:b}
     \end{subfigure}
     \hfill
     \begin{subfigure}{0.325\textwidth}
         \centering
         \includegraphics[width=\textwidth]{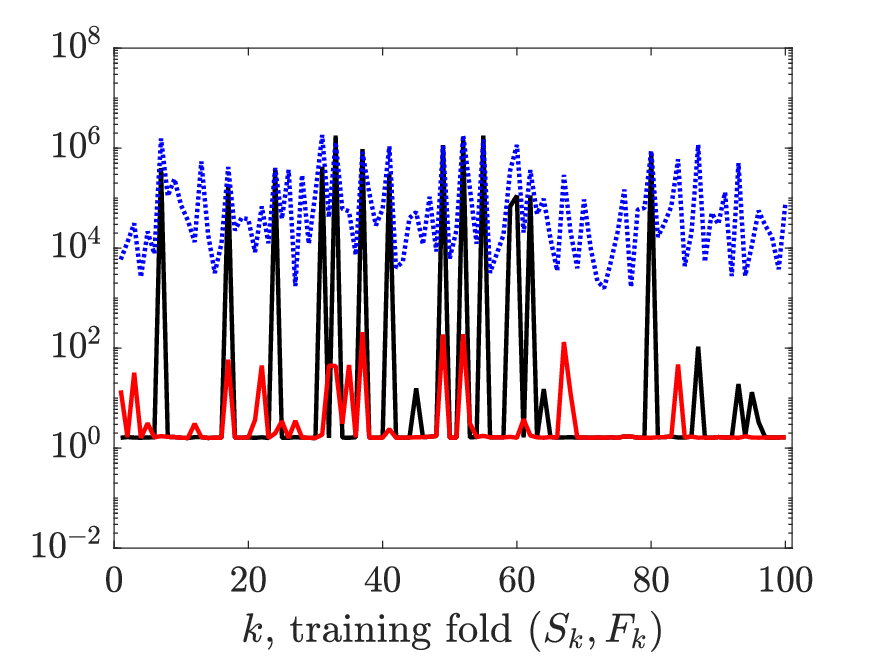}
         \caption{$\sigma_{\mathbf{S}} = 10^{-2}$ and $\sigma_{\mathbf{Y}} = 1$}
         \label{fig:c}
     \end{subfigure}
        \caption{{Comparison in terms of the cost $\{J_{k}\}_{k=1}^{100}$, given in Eq.~\eqref{eq:cost}, between \eqref{eq:op-side-info} and SINDy-SI for the identification of the Lorenz system~\eqref{eq:lorenz-ode} over  different noise levels. Noise is injected according to Eq.~\eqref{eq: data noise injection}.}}
        \label{fig:lorenz-deg5} \vspace{-25pt}
\end{figure*}

\begin{table}[htb]
    \centering
    \caption{Average performance of the system identification methods in terms of $\{J_{k}\}_{k=1}^{100}$ for the Lorenz system in Eq. \eqref{eq:lorenz-ode}. Noise is introduced in trajectory and output data with $\sigma_{\mathbf{S}}$ and $\sigma_{\mathbf{Y}}$, respectively. Best values are highlighted in bold text.} \vspace{-10pt}
    \begin{tabular}{|c||*{4}{c|}}
        \multicolumn{5}{c}{$\sigma_{\mathbf{S}} = 10^{-4}$ and $\sigma_{\mathbf{Y}} = 10^{-2}$}\rule[-0.75ex]{0pt}{3ex}\\\hline
        \makebox[2em]{Method}&\makebox[2em]{\eqref{eq:op-ols}}&\makebox[3em]{SINDy}&\makebox[3em]{\eqref{eq:op-side-info}}
        &\makebox[4em]{SINDy-SI}\rule[-0.75ex]{0pt}{3ex}\\\hline\hline
        Average $J_{k}$ &$5.63 \cdot 10^{4}$&$3.19 \cdot 10^{6}$&$822.42$&$\mathbf{2.0432}$\rule[-0.75ex]{0pt}{3ex}\\\hline
        \multicolumn{5}{c}{$\sigma_{\mathbf{S}} = 10^{-3}$ and $\sigma_{\mathbf{Y}} = 10^{-1}$}\rule[-0.75ex]{0pt}{3ex}\\\hline
        \makebox[2em]{Method}&\makebox[2em]{\eqref{eq:op-ols}}&\makebox[3em]{SINDy}&\makebox[3em]{\eqref{eq:op-side-info}}
        &\makebox[4em]{SINDy-SI}\rule[-0.75ex]{0pt}{3ex}\\\hline\hline
        Average $J_{k}$ &$5.59 \cdot 10^{5}$&$1.28 \cdot 10^{6}$&$1.89 \cdot 10^{4}$&$\mathbf{1.66 \cdot 10^{3}}$\rule[-0.75ex]{0pt}{3ex}\\\hline
        \multicolumn{5}{c}{$\sigma_{\mathbf{S}} = 10^{-2}$ and $\sigma_{\mathbf{Y}} = 1$}\rule[-0.75ex]{0pt}{3ex}\\\hline
        \makebox[2em]{Method}&\makebox[2em]{\eqref{eq:op-ols}}&\makebox[3em]{SINDy}&\makebox[3em]{\eqref{eq:op-side-info}}
        &\makebox[4em]{SINDy-SI}\rule[-0.75ex]{0pt}{3ex}\\\hline\hline
        Average $J_{k}$ &$5.13 \cdot 10^{6}$&$1.49 \cdot 10^{7}$&$2.16 \cdot 10^{5}$&$\mathbf{1.15 \cdot 10^{5}}$\rule[-0.75ex]{0pt}{3ex}\\\hline
    \end{tabular}
    
    \label{tab:ex1}%\vspace{-0.25cm}
\end{table}
\end{ex}
\begin{ex}[The SMIB system]
{Let us now assess the performance of SINDy-SI at identifying nonpolynomial dynamics.}
Consider the Single-Machine Infinite-Bus (SMIB) system \cite{smib}: \vspace{-0.2cm}
\begin{equation}
    \begin{split}
        \dot{\mathbf{x}} = \mathbf{f}(\mathbf{x}) = \begin{bmatrix}
            x_{2}\\
            \dfrac{1}{2 \alpha}\left(p_{m} - \dfrac{e' v}{\chi} \sin(x_{1}) - \beta x_{2}\right)
        \end{bmatrix}.\\
    \end{split}
    \label{eq:ode-smib}
\end{equation}
{
%where $x_{1}$ is the machine rotor angle, $x_{2}$ is the machine rotor speed deviation, $\alpha = 0.0106 ~s^{2}/rad$ the machine rotor inertia, $p_{m} = 1~pu$ the mechanical input power, $e' = 1.21~pu$ is the machine internal voltage, $v = 1~pu$ the voltage at the infinite bus, $\chi = 0.28~pu$ the total reactance between the machine internal node and the infinite bus, and $\beta = 0.03$ the machine damping. 
where $\alpha = 0.0106 $, $p_{m} = 1$, $e' = 1.21$, $v = 1$, $\chi = 0.28$ and $\beta = 0.03$. {Data were generated from $m = 100$ simulated trajectories of ODEs~\eqref{eq:ode-smib}, each with $r = 20$ timestamps linearly separated in the interval $[0,1]$. Noise, $\boldsymbol{\epsilon} \in \mathcal{N}(\mathbf{0},10^{-5},\mathbb{R}^{m r,n})$ and $\boldsymbol{\eta} \in \mathcal{N}(\mathbf{0},10^{-3},\mathbb{R}^{m r,n})$, is injected in the data set $D$ according to Eq.~\eqref{eq: data noise injection}.}

 The following side information was used in our implementation of SINDy-SI: 1) Equilibrium point at $ \mathbf{x}_{eq} = \left[\pi-\arcsin\left(\frac{p_{m} \chi}{e' v}\right)~0\right]^{\top}$, $S_{1} = \{\mathbf{f} \in \mathcal{C}^{1}(\mathbb{R}^{n};\mathbb{R}^{n}): \mathbf{f}(\mathbf{x}_{eq}) = \mathbf{0}\}$; 2) Local monotonic increase of the first coordinate $S_{2} = \{\mathbf{f} \in \mathcal{C}^{1}(\mathbb{R}^{n};\mathbb{R}^{n}): f_{1}(\mathbf{x}) \geq 0, x_{2} \geq 0\}$; 3) Local monotonic decrease of the first coordinate $S_{3} = \{\mathbf{f} \in \mathcal{C}^{1}(\mathbb{R}^{n};\mathbb{R}^{n}): f_{1}(\mathbf{x}) \leq 0, x_{2} \leq 0\}$. Side information 2 and 3 can be cast as SOS constraints $\{\hat{f}_{1}(\mathbf{x}) - z_{1}(\mathbf{x}) g(\mathbf{x}) - w_{1}(\mathbf{x}) x_{2}\}_{i=1}^{n} \in \Sigma\left(\mathbb{R}^{n};\mathbb{R}\right)$ and $\{-\hat{f}_{1}(\mathbf{x}) - z_{2}(\mathbf{x}) g(\mathbf{x}) + w_{2}(\mathbf{x}) x_{2}\}_{i=1}^{n} \in \Sigma\left(\mathbb{R}^{n};\mathbb{R}\right)$, respectively, with SOS multipliers $\{z_{i}(\mathbf{x}),w_{i}(\mathbf{x})\}_{i=1}^{2} \in \Sigma\left(\mathbb{R}^{n};\mathbb{R}\right)$ of appropriate degree and computational region defined in this example as $g(x)=25^2-\Vert \mathbf{x} \Vert_{2}^2$. 

{We used SINDy-SI without $\ell_{1}$ regularization ($\xi_2=0$) to fit a degree 7 model  with $h = 36$ monomials ($h > r$) to the trajectory data with sparseness threshold $\lambda = 10^{-2}$. {The best fitting trained model obtained from cross-validation was found to be $\hat{f}_{1}(\mathbf{x})=0.9999 x_2$ and $\hat{f}_{2}(\mathbf{x})=0.1790 x_1^7 - 1.944 x_1^5 + 34.09 x_1^3 - 203.8538 x_1 - 1.415 x_2 + 47.171$.}
% The average model obtained from the mean value of the coefficients associated with the trained models over each data fold was found to be $\hat{f}_{1}(\mathbf{x})=0.9999 x_2$ and $\hat{f}_{2}(\mathbf{x})=0.1790 x_1^7 - 1.944 x_1^5 + 34.09 x_1^3 - 203.8538 x_1 - 1.415 x_2 + 47.171$.
%
%\begin{equation}
%    \scriptstyle \hat{\mathbf{f}}_{\text{avg}}(\mathbf{x}) = \begin{bmatrix}\scriptstyle
 %       0.9999 x_2\\
  %      \scriptstyle 0.1790 x_1^7 - 1.944 x_1^5 + 34.09 x_1^3 - 203.8538 x_1 - 1.415 x_2 + 47.171
   % \end{bmatrix}
%\end{equation}}
Interestingly, in this numerical example we witnessed some emergent behavior of the SINDy-SI algorithm. Unexpectedly, the learned SINDy-SI model closely approximated the $7$'th order Taylor expansion of the SMIB system~\eqref{eq:ode-smib}, $f_1(\mathbf{x})=x_2$ and $f_2(\mathbf{x}) \approx 0.0404 x_1^7 - 1.6987 x_1^5 + 33.97 x_1^3 - 203.841 x_1 - 1.415 x_2 + 47.1698$.
%\begin{equation}
%    \mathbf{f}(\mathbf{x}) \approx \begin{bmatrix}
%        x_2\\
%       0.0404 x_1^7 - 1.6987 x_1^5 + 33.97 x_1^3 - 203.841 x_1 - 1.415 x_2 + 47.1698
%    \end{bmatrix}
%\end{equation}
}
}
\end{ex}
\vspace{-0.2cm}
\section{Conclusion}\label{sec:conclusions}

This paper introduces the SINDy-SI approach to system identification, which combines sparsity techniques with side information constraints. This innovative approach strikes balance between model accuracy and generalization, allowing it to effectively capture the underlying governing dynamics without succumbing to overfitting. {Through numerical simulations, we demonstrated that SINDy-SI, as a result of the combination of side information constraints and sparseness, outperforms both traditional SINDy and the SOS-based approach presented in \cite{ahmadi2023} and yields better models in the case of scarce noisy data where SINDy generally fails.}

\bibliographystyle{ieeetr}
\bibliography{bibliography}

\end{document}